\documentclass[fleqn,usenatbib]{mnras}

\usepackage{newtxtext,newtxmath}

\usepackage[T1]{fontenc}

\DeclareRobustCommand{\VAN}[3]{#2}
\let\VANthebibliography\thebibliography
\def\thebibliography{\DeclareRobustCommand{\VAN}[3]{##3}\VANthebibliography}

\usepackage{graphicx}	
\usepackage{amsmath}	
\usepackage{lipsum}

\newcommand{\hst}{\textit{HST}}
\newcommand{\jwst}{\textit{JWST}}
\newcommand{\rst}{\textit{Roman}}
\newcommand{\angstrom}{\mbox{\normalfont\AA}}

\def\flamunit{\ifmmode {\rm\,erg\,cm^{-2}\,s^{-1}\,\angstrom^{-1}}\else
                ${\rm\,ergs\,cm^{-2}\,s^{-1}\, \angstrom^{-1}}$\fi}
\def\oIII{[OIII]}
\def\oII{[OII]}
\def\nV{[NV]}
\def\sII{[SII]}
\def\nII{[NII]}

\title[{ESpRESSO}]{ESpRESSO -- Forward modeling \textit{Roman Space Telescope} spectroscopy}

\author[Gabrielpillai et al.]{Austen Gabrielpillai$^{1, 2, 3}$\thanks{E-mail: a.gabrielpillai@gmail.com},
Isak G. B. Wold$^{1, 2, 3}$, 
Sangeeta Malhotra$^{2}$, 
James Rhoads$^{2}$, 
Guangjun Gao$^{2}$,\newauthor 
Mainak Singha$^{1, 2, 3}$,
and Anton M. Koekemoer$^{4}$
\\
$^{1}$Institute for Astrophysics and Computational Sciences, Catholic University of America, 20064, USA\\
$^{2}$Astrophysics Science Division, Goddard Space Flight Center, Greenbelt, MD 20771, USA\\
$^{3}$Center for Research and Exploration in Space Science and Technology, NASA/GSFC, Greenbelt, MD 20771, USA\\
$^{4}$Space Telescope Science Institute, 3700 San Martin Drive, Baltimore, MD 21218, USA
}

\date{Accepted XXX. Received YYY; in original form ZZZ}

\pubyear{\the\year{}}

\begin{document}
\label{firstpage}
\pagerange{\pageref{firstpage}--\pageref{lastpage}}
\maketitle

\begin{abstract}
We describe the software package \texttt{ESpRESSO} - [E]xtragalactic [Sp]ectroscopic [R]oman [E]mulator and [S]imulator of [S]ynthetic [O]bjects, created to emulate the slitless spectroscopic observing modes of the {\it Nancy Grace Roman Space Telescope} (\rst) Wide Field Instrument (WFI).  We combine archival \textit{Hubble Space Telescope} (\hst) imaging data of comparable spatial resolution with model spectral energy distributions to create a data-cube of flux density as a function of position and wavelength. This data-cube is used for simulating a nine detector grism observation, producing a crowded background scene which model field angle dependent optical distortions expected for the grism. We also demonstrate the ability to inject custom sources using the described tools and pipelines. In addition, we show that spectral features such as emission line pairs are unlikely to be mistaken as off order contaminating features and vice versa.  Our result is a simulation suite of half of the eighteen detector array, with a realistic background scene and injected Ly$\alpha$ emitter (LAE) galaxies, realized at 25 position angles (PAs), 12 with analogous positive and negative dithers, Using an exposure time of 10ks per PA, the full PA set can be used as a mock deep \rst\ grism survey with high (synthetic) LAE completeness for developing future spectral data analysis tools. 
\end{abstract}



\section{Introduction}
A galaxy's spectrum contains a wealth of information about its contents and physical conditions.  The continuum, absorption lines, and emission lines tell a coupled story about stellar population ages, star formation, nuclear activity (i.e., AGN), dust content, and chemical abundances, as well as redshift and distance measures  \citep{calzetti:1994, calzetti:2001, fiore:2012, leja:2019, papovich:2022}. Even the ionization state of the intergalactic medium (IGM) between the observer and source can be inferred via spectral feature analysis \citep{gunn:1965, becker:2001, mcdonald:2006}.

Combining spectral and direct imaging data has diverse astrophysical applications spanning multiple scales. The Extended Baryon Oscillation Spectroscopic Survey (eBOSS) \citep{dawson:2016:sdss:eboss}, a sub-program of the Sloan Digital Sky Survey (SDSS) \citep{abazajian:sdss:vii, blanton:2017:sdss}, was instrumental in mapping the local universe ($0.6 < z < 1.1$), cataloging populations of luminous red galaxies (LRGs) and emission line galaxies (ELGs) to constrain measurements of the expansion of the Universe due to Dark Energy. SDSS spectral information combined with crowd-sourced morphological classifications \citep{lintott:2008:zoo} led to discovery of the Green Pea galaxies \citep{cardamone:2009, jaskot:oey:2013}, low metallicity dwarf galaxies with high star formation rates that live low density environments. The \textit{James Webb Space Telescope} (\jwst) \citep{gardner:2006:jwst, gardner:2023:jwst, rigby:2023:jwst}  has provided a wealth of spectroscopic observations of objects in the early universe, including Green Pea galaxy analogs \citep{rhoads:2023} at $z \sim 8$, high-$z$ candidates with incredibly high star formation efficiency \citep{bunker:2023, harikane:2024}, and multiple contenders of the earliest galaxies \citep{arrabal:2023, cl:2023, harikane:2023}.

Galaxy spectra are generally collected using slit spectrographs, fiber spectrographs, integral field units, or slitless spectroscopy.  
Among these, slitless spectroscopy is unique in yielding spectra of every object within the survey field, with no preselection of particular objects or regions of the focal plane where light will be dispersed.  This yields a unique benefit in survey completeness, along with unique data analysis challenges posed by overlap between spectra of different objects.  Slitless spectroscopy elements include prisms and grisms, the key difference being that grisms allow an undeviated wavelength to pass straight through to the detector.  

Slitless spectroscopy from space is a particularly powerful approach for studying high-redshift galaxies.  These objects are compact, with angular sizes ($\sim 0.1\arcsec$) comparable to space-based imaging resolution.  Moreover, their rest-UV light is redshifted to near-infrared wavelengths, where Earth's atmosphere glows brightly.  Space telescopes thus win doubly--- both on the surface brightness of foreground sky emission, and on the area of sky emission that contributes noise to each spectrum.
This has led to a long history of slitless spectroscopy with \textit{Hubble Space Telescope} (\hst), including the GRAPES \citep{Malhotra2005,Pirzkal2004},
PEARS \citep{Straughn2008,Rhoads2009,Xia2012},
3D-HST \citep{3d-hst:brammer,momcheva:2016},
and FIGS \citep{Tilvi2016,Pirzkal2017,Ferreras2019,Pharo2020} surveys.
\textit{JWST} has implemented further slitless spectroscopic capabilities both on its NIRCam and NIRISS instruments \citep{matthee:2023, sun:2023,wang:2023, kashino:2023, eisenstein:2023, meyer:2024}.

The \textit{Nancy Grace Roman Space Telescope} (\rst) \citep{spergel:2015:wfirst, akeson:2019:roman}, initially known as Wide-Field Infrared Survey Telescope (WFIRST), is NASA's next up-and-coming flagship mission. Slated to launch by mid-2027, the telescope will have a field-of-view (FOV) of $0.281 \ {\rm deg}^2$ across 18 detectors, approximately $100\times$ and $200\times$ \hst's optical and NIR sky coverage respectively. Onboard is the Wide Field Instrument (WFI) \citep{pasquale:2018}, which is comprised of 8 filters spanning the optical to near-infrared (NIR) ($0.28 - 2.3\ \micron$), as well as grism ($1.0 - 1.93\ \micron$) and prism ($0.75 - 1.80\ \micron$) slitless spectroscopy elements. Simulations of \rst's optical elements have predicted applications to galaxy clustering statistics \citep{yung:roman, perez:2022}; constraints on the faint end quasar luminosity function \citep{tee:2023}; Dark Energy experiments via weak-lensing \citep{troxel:2021}; and core-community survey planning \citep{dore:2018, wang:2022}. WFI will also be a key instrument in detecting $z \geq 7$ Lyman-$\alpha$ emitters (LAEs) \citep{wold:2023}, galaxies with a bright Ly$\alpha$ emission feature that is only visible in the ionzied IGM, making them incredible probes of cosmic reionization \citep{Malhotra2004,dijkstra:2014}. 

The science that will be enabled by \rst\ will be monumental in the field of galaxy formation and evolution \citep{somerville:dave:2015, newman:gruen:2022}. However, reducing such a treasure trove of information comes with its own challenges. Highly clustered galaxies can blend together and contaminate each other both in direct and spectral imaging, making it difficult to extract and recover fainter sources such as high-$z$ galaxies. Spectral features can also be confused with contamination features. There are also data processing challenges. While simulations of \rst's spectroscopic capabilities do exist \citep{wold:2023}, the existing software for producing synthetic observations \citep{axe:kummel} was originally developed for the  \hst\ and had to be modified to best mimic \rst's optics with various trade offs. It is important to test future spectral extraction software on high-fidelity, accurate images, presenting the need to develop \rst-specific simulation software that is highly modular. 

In this work, we present our \rst\ grism simulation code, \texttt{ESpRESSO}, built explicitly to accurately model various optical and detector effects in WFI slitless spectroscopic modes. This work is the first step in modeling WFI spectroscopy including optical distortions to high precision, laying the groundwork for future features, additions, and modularity while remaining accessible. We expect to release our simulated grism scenes to support development and testing of grism data analysis tools. Our data release will primarily consist of a foreground simulated at 25 position angles (PAs), akin to a deep \rst\ grism survey with 10 ks exposure time per PA \citep{wold:2023}.

This paper has the following structure. We describe the high resolution \hst\ images of the COSMOS field from the CANDELS program and best-fit spectral energy distributions (SEDs) from 3D-HST used as inputs to our simulation (\S\ref{sec:data}). We produce pipelines to create high resolution simulated noiseless grism foreground images with multiple orders at different orientations; add photon noise according to the integration time; and inject custom source objects into our images (\S\ref{sec:grism}). Finally, we compare our simulations to other relevant state-of-the-art images produced as part of \rst\ preparatory work, discussing contamination effects, caveats, and next steps for our own pipeline (\S\ref{sec:discussion}).

\section{Data}
\label{sec:data}
In this section, we outline the different images and catalogs used in the production of our slitless spectroscopic simulations. In Sec.~\ref{sec:data:candels&cosmos} we cover the CANDELS project and one of the observed fields, COSMOS. In Sec.~\ref{sec:data:3dhst&eazy}, we describe the 3D-HST Treasury program and the different software (e.g. EAZY) previously utilized to produce the foreground spectra used in this work. We also discuss basic data reduction and cleaning strategies done as part of our pre-processing. 

\subsection{CANDELS \& The COSMOS Field}
\label{sec:data:candels&cosmos}

The Cosmic Assembly Near-IR Deep Extragalactic Legacy Survey (CANDELS) \citep{CANDELS:grogin, CANDELS:koekemoer} was a legacy project that utilized \hst's Wide Field Camera 3 (WFC3) and Advanced Camera for Surveys (ACS) to conduct both wide and deep field surveys of five fields (GOODS-North, GOODS-South, EGS, UDS, and COSMOS) in the near-IR range. ``CANDELS/Wide'' imaged all five fields with J-band (WFC3/IR F125W) and H-band (WFC3/IR F160W) filters to reach a depth of 27.0 mag per filter, the purpose was identify bright high-z galaxy candidates for follow up observations. ``CANDELS/Deep'' imaged only two fields (GOODS-North and -South) and added observations with the Y-band filter (WFC3/IR F105W), reaching an H-band depth of 27.7 mag. Two of the primary survey objectives included studying ``cosmic dawn" ($z \gtrsim 3$) and ``cosmic high noon" ($1.5 < z < 3$).  Cosmic dawn studies included identifying galaxies at the end of the Epoch of Reionization and their observable properties (e.g. star formation rates, stellar mass, etc.) plus verifying high-redshift active galactic nucleus (AGN) candidates to constrain the AGN luminosity function for $z > 6-7$. Cosmic high noon observation studied the peak of cosmic star formation \citep{madau:2014} and AGN activity by using broad-band spectral energy distributions (SEDs) to measure redshifts, star formation rate, and stellar mass for galaxies down to $m_{*} = 2 \times 10^{9} \ {\rm M_\odot}$, as well as by identifying rest-frame morphology and substructure. 

\begin{figure}
	\includegraphics[width=0.995\columnwidth]{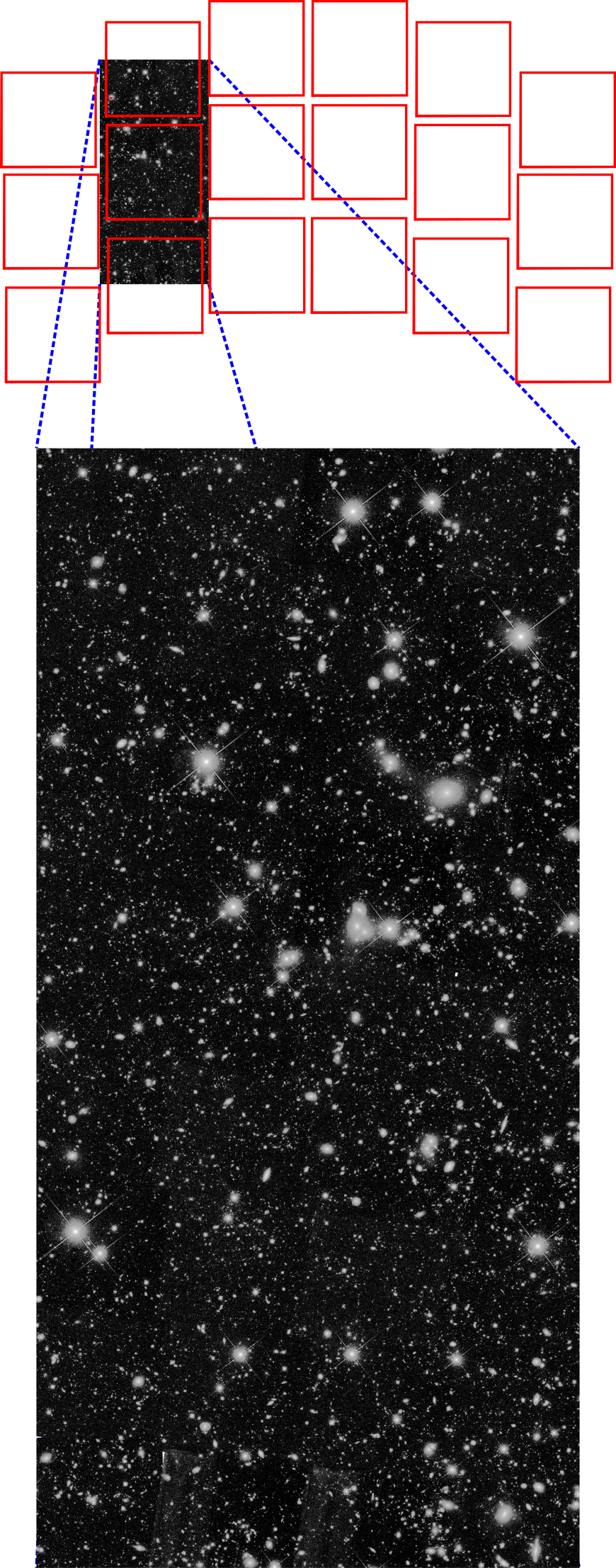}
    \caption{Top: \rst\ detector array overlaid on top of our science image input. Bottom: Blown up version of the science image input. Specifically shown is the F160W cutout of the CANDELS COSMOS field at a resolution of 30 mas / pixel. The area of the image is 15' x 6' or 0.025 deg$^2$.}
    \label{fig:sci-img-cutout}
\end{figure}

The Cosmic Evolution Survey (COSMOS) field \citep{cosmos:scoville, cosmos:koekemoer} is a historic and continuously well observed area. The field was designed to survey galaxy formation and evolution from $0.5 < z < 6$ while probing for dark matter, large scale structure, and AGN and is easily accessible by both ground- and space-observing facilities for multi-band imaging (from radio to x-ray). The original \hst\ survey was conducted with the ACS/WFC F814W (broad I-band) filter at a depth of 27.2 mag over a $\sim 2$ deg$^{2}$ area, part of which was later selected as one of the ``CANDELS/Wide" fields. We chose this CANDELS/Wide COSMOS region as the basis of our simulations because it offers a key combination of detailed morphological information and SEDs over a relatively wide field of view.
We will primarily use F160W image products in this work as not only is the filter's central wavelength (1.60 \micron)  close to \rst\ grism's undeviated wavelength (1.55 \micron), but it was the widest deep NIR actual imaging available.
Thus, we can reasonably assume that identified F160W object morphology will be adequate inputs in our image processing as outlined later. Fig.~\ref{fig:sci-img-cutout} showcases a 0.025 deg$^{2}$ cutout of the CANDELS F160W COSMOS field mosaic at a resolution of 30 milli-arcsecond per pixel (mas / pixel) \citep{CANDELS:koekemoer}, shown in comparison to \rst's detector array. 

\subsection{3D-HST \& EAZY}
\label{sec:data:3dhst&eazy}
3D-HST \citep{3d-hst:brammer} was a \hst\ Treasury program designed to complement the CANDELS program by utilizing WFC3 imaging and grism spectroscopy on the same five fields. One of the primary goals was to precisely measure  redshifts for a galaxy population between $1 < z < 3$ at an H-band magnitude up to approximately 23.8. This was accomplished through the combination of imaging with both the ACS and WFC3 grisms, covering both optical and near-infrared spectra. WFC3's G141 grism is fairly similar to \rst's in its wavelength coverage, ranging from $1.05 - 1.70 \mu m$. It is able to capture spectral emission lines such as H$\alpha$ ($0.7 < z < 1.5$), \oIII\ ($1.2 < z < 2.3$), and \oII\ ($2.0 < z < 3.4$). Direct imaging with the F140W filter is used to calibrate and extract spectra from the grism observations. A segmentation map was first derived from the F140W mosaic using the Source Extractor software \citep{sextractor:bertin&arnouts}. This was then used together with the G141 observations \citep{momcheva:2016} as input to the aXe software \citep{axe:pirzkal,axe:kummel}. This produced extracted 1D and 2D spectra for each object in the field.  

As part of the 3D-HST program, \cite{3d-hst:skelton} aggregated a photometric catalog for objects in the COSMOS field via joint image and data reduction of surveys conducted via the Canada France Hawaii Telescope \citep{erben:2009, hildebrandt:2009, bielby:2012}, Subaru Telescope \citep{taniguchi:2007}, Nicholas U. Mayall 4-meter Telescope \citep{whitaker:2011}, Visible and Infrared Survey Telescope for Astronomy \citep{mccracken:2012}, Spitzer Space Telescope \citep{sanders:2007, ashby:2013}, and \hst\ \citep{CANDELS:grogin, CANDELS:koekemoer, 3d-hst:brammer}. All fluxes were normalized to an AB zero point of 25. The photometric catalog and 44 filter profiles were used as input in EAZY (Easy and Accurate Redshifts from Yale) redshift code \citep{brammer:eazy} to create both the best-fit SEDs and photometric redshifts used in this work. Segmentation maps--- i.e., images identifying which pixels belong to which object and mapping them to the catalog---  were also generated in this work for CANDELS F125W and F160W mosaics using lower-resolution maps (60 mas / pixel) as a base \citep{3d-hst:skelton}. We note that the catalog IDs map both to the SEDs and object IDs in the segmentation map, linking all three data products in our pipeline.

\begin{figure}
    \includegraphics[width=\columnwidth]{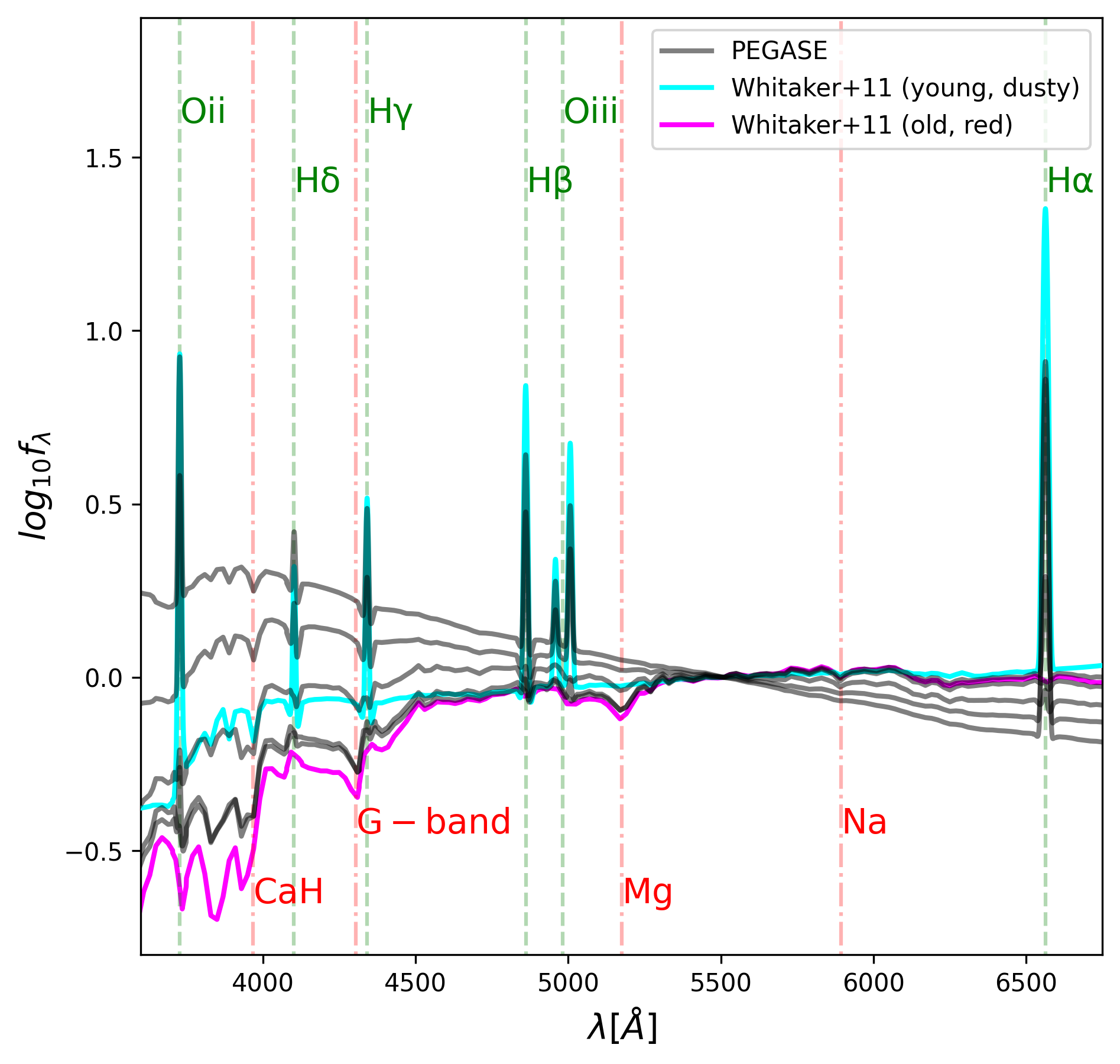}
    \caption{EAZY templates used to re-construct best fit SEDs in the rest frame. Templates 1--5 (black) are from the PEGASE model \citep{fioc:rocca:1997}, template 6 (cyan) and 7 (magenta) are derived from observations in \citet{whitaker:2011}. Absorption (red, dashed-dotted) and emission (dashed) line features in the templates are annotated. } 
    \label{fig:eazy-templates}
\end{figure}

EAZY \citep{brammer:eazy} is software used for producing photometric redshifts and best-fit synthetic SEDs. The process involves inputting a catalog of observed photometric fluxes, where each object is fit to a linear combination of spectral templates that pass through the input fluxes. \cite{3d-hst:skelton} use seven templates to fit the aforementioned 44 filter profiles - five from the PEGASE spectral evolution model with emission lines \citep{fioc:rocca:1997}; a young, dusty galaxy template; and an old, red galaxy template \citep{whitaker:2011}. The best-fit weight per template is stored, allowing for quick and rapid reconstruction of the synthetic SED. Fig. \ref{fig:eazy-templates} depicts the seven templates used to reconstruct each best-fit SED, where they are linearly combined with continuum attenuation, returning an object's observed frame spectra in units of $10^{-17}$ erg / s / cm$^2$ / \angstrom. We refer the reader to \cite{brammer:eazy} for a more in-depth description of the linear combination process. Although some templates contain a Ly-$\alpha$ emission line feature, none of the objects in the catalog are at a high enough redshift for the emission line to fall in \textit{Roman}'s grism spectral range of 1.0--1.93 $\mu$m. Other notable emission lines that are both present in the templates and resolvable by the grism  are H$\alpha$, \oIII, and \oII, as seen in in Fig.~\ref{fig:eazy-templates}.  After SED fitting, each galaxy's emission line properties (and classification as a line emitter if appropriate) can be determined by calculating line fluxes after subtracting the continuum flux estimated from adjacent wavelengths.

\subsubsection{Correcting point source and extended SEDs}
\label{sec:sed_correction}

EAZY as implemented by \cite{3d-hst:skelton} does not include stellar templates.   It also failed to find valid fits for extremely bright and extended galaxies, effectively failing to produce a realistic SED for these types of objects. 
It is important to model these objects correctly to have a realistic foreground scene, as they act as contaminants to surrounding galaxies. We calculate the expected F160W magnitude of each object as a step to classify which objects are not being represented correctly. We follow the process as outlined in \cite{vacca&tokunaga:2005} to calculate our source magnitude for a given filter using our EAZY spectra. We first calculate the effective (pivot) wavelength of the F160W filter using its wavelength response, $R_{\rm F160W}(\lambda)$: 

\begin{equation}
\label{eq:lam_eff}
    \lambda_{\rm eff, F160W} = \sqrt{\frac{\int \lambda R_{\rm F160W}(\lambda) d\lambda }{\int \lambda^{-1} R_{\rm F160W}(\lambda) d\lambda}} \\
\end{equation}

The effective wavelength is subsequently used to convert an object's observed (EAZY) $F_{\lambda}$ to $F_{\nu}$, the flux per unit frequency (erg / s / cm$^2$ / Hz), for the specified bandpass as stated in \cite{vacca&tokunaga:2005}: 

\begin{equation}
\label{eq:f_nu}
    F_{\nu, {\rm obs}} = \frac{\lambda_{\rm eff, F160W}^{2}}{c} \frac{\int \lambda f_{\rm obs}(\lambda) R_{\rm F160W}(\lambda) d\lambda}{\int \lambda R_{\rm F160W}(\lambda) d\lambda}
\end{equation}

We note that $F_{\nu, {\rm obs}}$ has units of flux per unit frequency, while $f_{\rm obs}$ has units of flux per unit wavelength. 
The integrals in Eq. \ref{eq:lam_eff} and \ref{eq:f_nu} are performed over the filter's wavelength range of 1.38 - 1.70 $\mu$m. We last use the following to calculate the object's observed AB magnitude as given in \cite{oke&gunn:1983}:

\begin{equation}
    {m}_{\rm F160W}^{\rm obs} = -2.5 {\rm log}_{10} (F_{\nu, \rm obs}) - 48.6
\end{equation}

Our observed magnitude is then cross compared with the cataloged F160W magnitude. Within our COSMOS field of view, EAZY's $\chi^2$ fit fails to converge for 400 bright stars (3D-HST star flag$=1$, $m_{\rm F160W}<23$ and $\chi^2>1000$) and 34 nearby extended galaxies ($m_{\rm F160W}<23$ and abs($m^{\rm obs}_{\rm F160W}-m^{\rm cat}_{\rm F160W})>0.8$).  To reproduce an extra-galactic foreground scene, we need to recover accurate infrared spectra for these bright stars and galaxies, and we adopt the following procedures to model these objects.

For stars, we fit their cataloged 3D-HST F606W, F125W, and F160W fluxes \citep{3d-hst:skelton} to a grid of Kurucz (1993) ATLAS model stellar atmospheres that span spectral types O through M.  We adopt the re-normalized model with the best $\chi^2$ value as our input stellar spectrum. For extended galaxies, we fit a $\chi^2$ power-law to their 3D-HST F606W, F125W, and F160W fluxes. Following these methods, we produce replacement infrared spectra for these bright stars and galaxies that are not well modeled by EAZY. Note that the replacement spectra overwrite the failed spectra in the SED library.

\section{Grism Simulations with \texttt{ESpRESSO}}
\label{sec:grism}

\begin{figure*}
    \includegraphics[width=\textwidth]{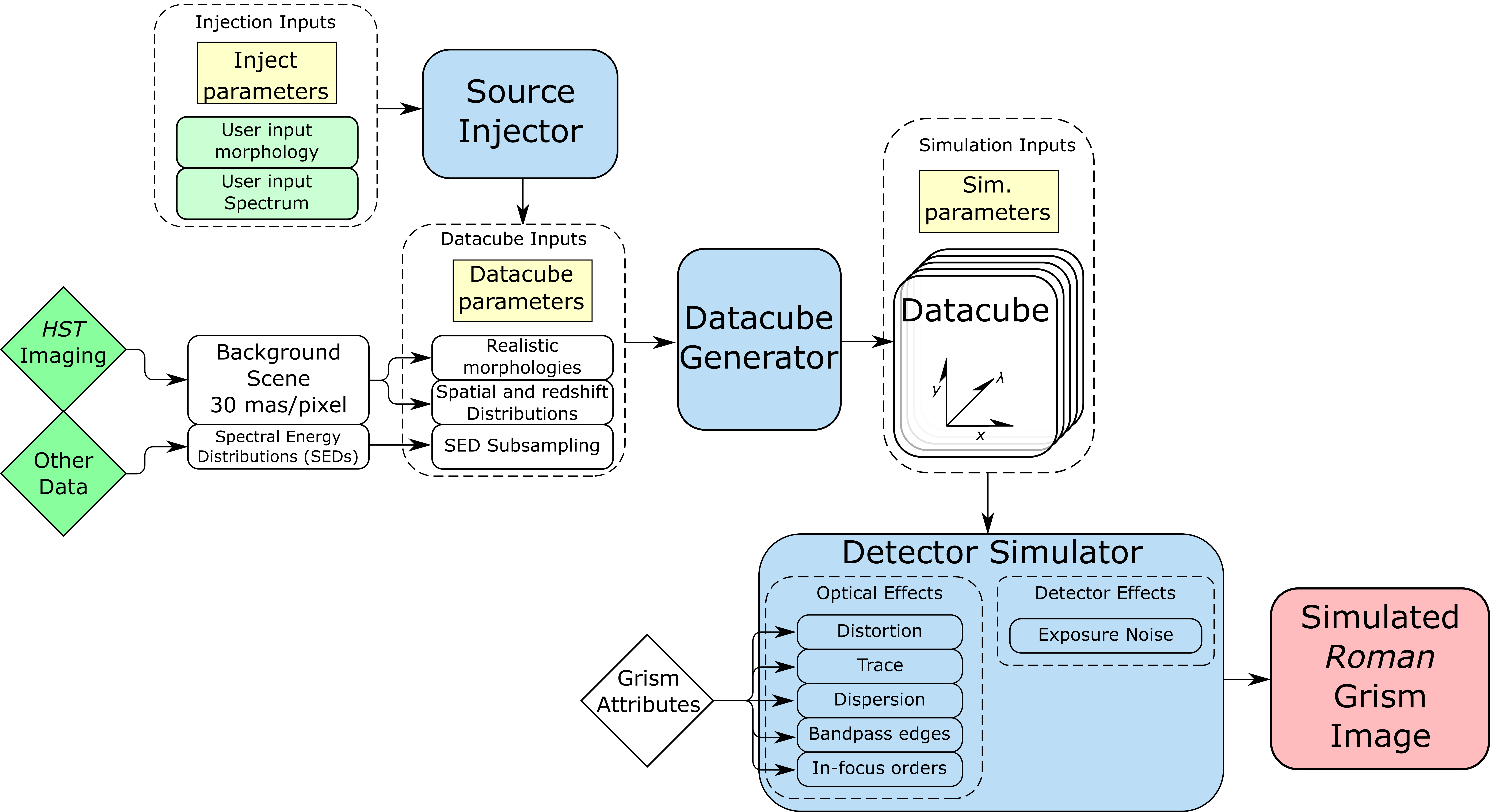}    
    \caption{Schematic of the \texttt{ESpRESSO} pipeline. The algorithms and methods developed in this work are blue rounded squares. Using our imaging data, we generate a datacube ($\S$\ref{sec:grism:data-cube}) that is a function of (x, y, $\lambda$). This datacube is used as input in our grism detector simulator ($\S$\ref{sec:grism:polynomial}-\ref{sec:grism:flux-assign}), producing a synthetic foreground image. Users can also generate grism images of their custom objects ($\S$\ref{sec:grism:source-injection}) using the same pipeline. Input related data products are in green and final output is in red.} 
    \label{fig:schematic}
\end{figure*}

\rst's grism instrument spans a wavelength range of 1.00--1.93 $\micron$ with an undeviated wavelength of 1.55 $\micron$. We perform a series of image processing and transformation techniques to produce mock slitless spectroscopic foreground images from the image data cube.
We outline the major steps in our pipeline to produce our synthetic grism images. A visual representation of our pipeline can be seen in Fig. \ref{fig:schematic}, providing a brief highlight on the inputs and the different computational processes utilized in this work.
We first start by constructing an image datacube as a function of 2D position and wavelength (Sec. \ref{sec:grism:data-cube}). We then describe the equations used to convert from sky coordinates to grism coordinates (Sec. \ref{sec:grism:polynomial}) as well as the distortions involved in flux assignment (Sec. \ref{sec:grism:flux-assign}). We follow up with our post-processing steps, such as injecting custom sources into our foreground images (Sec. \ref{sec:grism:source-injection}) and modeling noise (Sec. \ref{sec:grism:additional}). Last, we discuss the flexibility of our simulation with parameter file inputs (Sec. \ref{sec:grism:parameters})

\subsection{Image Cube Construction}
\label{sec:grism:data-cube}
The image datacube, produced as a function of pixel position and wavelength (x, y, $\lambda$), will be used as the primary input in the grism simulation step. We choose to store and save the datacube rather than reconstructing it per simulation run to avoid a substantial increase in run time and computation costs. This simplifies the process for creating multi-detector simulations at different position angles, where the only transform needed is a rotation around a specified axis. A science image, segmentation map, an SED library, an object catalog, and a parameter file are needed for data cube construction.  The inputs used in this work have been previously outlined in Sec. ~\ref{sec:data}. 

Our input science image, presented in Fig. \ref{fig:sci-img-cutout}, has been stretched in logarithmic scaling in order to bring out fainter foreground objects into view, emphasizing the number of objects crowding the field. Notably included are bright and extended stars and galaxies acting as contaminants when dispersing spectra. A world coordinate system (WCS) is attached in the header of the image, allowing for transformation from pixel grid coordinates to sky coordinates while maintaining image distortions. As mentioned, the F160W segmentation map was generated from a lower resolution F125W image (60 mas / pixel), key for extracting and maintaining object morphology from the science image. These two types of images are used in conjunction to first create our 'noiseless' datacube as a function of wavelength via each object's SED. 

We first preform a series of image manipulation processes on our two images. If specified in the parameter file, the segmentation map and science image are cropped to reduce the total imaging area and remove non-physical artifacts such as voids that result from missing data. Any objects that land on the cropping boundary are removed from the images to avoid clipped objects that will produce non-physical results. This is accomplished by identifying the affected object IDs in the segmentation map, subsequently setting those values in the map's image array to 0. The same pixels in the science image do not need to be set to 0 as those pixels are now excluded from the simulation based on the segmentation map. The segmentation map is used to extract source pixels and ignore as much background noise as possible. Any pixel that does not belong to an object in the segmentation map is set to a value of 0. Non-physical flux values in the science image such as pixels with negative fluxes are also set to 0. We choose to then represent our post processed image as a compressed sparse row (CSR) matrix, a data representation that stores the information of non-zero matrix values, effectively reducing the number of pixels to work with by a factor of 10. The area of our cutout used in this work can be again seen in Fig. \ref{fig:sci-img-cutout}.

The SED library contains the SEDs of all objects in the field and is stored in the HDF5 format. The original SED does not have a uniform $\lambda_{\rm step}$, so we linearly interpolate the 1D spectra ($f_{\rm obs, interp}$) to best approximate the observed flux at a given wavelength. The SEDs are re-sampled to have a wavelength interval of $\lambda_{\rm step}$ = $10^{-4}$ \micron\ (1 \angstrom\ ) to avoid resolution effects such as banding appearing in the grism simulations. We then calculate $f_{\rm obs}(\lambda + \frac{\lambda_{\rm step}}{2})$, the discrete spectrum flux density evaluated at the midpoint between $(\lambda, \lambda + \lambda_{\rm step})$, via the following: 

\begin{equation}
    f_{\rm obs}(\lambda + \frac{\lambda_{\rm step}}{2}) = \frac{1}{\lambda_{\rm step}} \int_{\lambda}^{\lambda + \lambda_{\rm step}} f_{\rm obs, interp} (\lambda) d\lambda
\end{equation}

where we evaluate the integral via the trapezoidal rule in order to preserve flux conservation. The resulting flux is transformed back into flux density by dividing out the wavelength step due to the response function converting from flux density to counts per second.  Each SED has been re-sampled to a wavelength interval of $\lambda_{\rm step}$ between 1.00--1.93 $\micron$ split into $0.93\lambda_{\rm step}^{-1}$ intervals. 

The source object pixels' image positions and respective values were previously extracted from the input image using the segmentation map, a process that retains the morphology of the object. The following equation describes the pixel flux density reassignment process, resulting in a monochromatic image of an object at a given $\lambda$:

\begin{equation}
    f_{\rm obs, px}(\lambda) = f_{\rm obs, src}(\lambda) \frac{f_{\rm px, filt}} {\sum_{i=1}^{n_{\rm px}} f_{i, \rm filt}}
\end{equation}

Using each source's new, resampled observed SED, $f_{\rm obs, src}(\lambda)$, the observed flux density at $\lambda$ is multiplied by a pixel's individual fractional contribution to the source object, calculated by dividing its value ($f_{\rm px, filt}$) by the sum for all pixels that compose the source object. As mentioned, our choice of filter is F160W. The process is done for all objects in the field, producing an image between the wavelength range of ($\lambda$, $\lambda + \lambda_{\rm step}$). Once completed for all objects, each 'instantaneous' science image slice is stored as a CSR matrix rather than the standard n x m matrix representation.
 This procedure is repeated $0.93\lambda_{\rm step}^{-1}$ times for each wavelength in the wavelength array, producing the image datacube that will be used in the grism simulation step.

\subsection{Detector pixel assignment}
\label{sec:grism:polynomial}
The first step in forward modeling \rst's grism is the sky-to-grism detector coordinate transformation. The original science image's WCS is used to transform from the science image pixel coordinates to sky coordinates, which maintains inherent image distortions. The science image pixel coordinates need to be shifted to their original values if the image has been cropped in the data cube creation step (as noted in the parameter file) to preserve WCS transformations. The midpoint of the data cube image is assigned as field center which can be treated as a local origin center, where a rotational transformation is then applied to all points to simulate the instrument pointing a non-zero position angle. We are simulating multiple angles as a means to separate the 2D spectra of two nearby sources along the dispersion axis, also known as source contamination. To simulate dithering, we translate the local image origin by a small $\delta x$ or $\delta y$ of 0.165" (1.5 \rst\ WFI pixels).

The grism coordinate vector $p_{\rm grism}$ is calculated via a 22-term 2D-fit polynomial (consisting of a total of 44 best fit coefficients) which takes in sky coordinates (degrees) and a wavelength (\micron) as input, resulting in a 2D positional grism coordinate vector (mm) that is subsequently transformed into detector (pixel) coordinates. Each detector has its own unique polynomial to transform from sky to grism coordinates, meaning that SCA\#5's polynomial cannot be used to assign fluxes to SCA\#3 and that each detector needs to be simulated independently. The polynomial is represented in the following simplified form:
\begin{gather}
    \overrightarrow{p}_{\rm grism}(x, y, \lambda)
   =
   \begin{bmatrix}
    c_{1, x} \ c_{2, x} \ \dots \ c_{22, x}\\ 
    c_{1, y} \ c_{2, y} \ \dots \ c_{22, y}\\ 
   \end{bmatrix}
   \begin{bmatrix}
    t_{1} \\ t_{2} \\ \vdots \\ t_{22}\\ 
   \end{bmatrix}
\end{gather}
where $t_{i}(x, y, \lambda) = x^{a} y^{b} \lambda^{c}$ and $a, b, c \in [0, 5]$. We note that $t = 0$ for all 4th power and nearly all 5th power terms. The result is a (x, y) position given in units of [mm]. Using the field center of the detector $p_{ \rm center} = p_{\rm grism}(0, 0, \lambda)$ and a detector (SCA) width of 4096 pixels, the detector position $\overrightarrow{p}_{\rm SCA}= \begin{bmatrix}p_{\rm SCA, x} \ p_{\rm SCA, y} \end{bmatrix}$ for the (1,1) grism order can be approximated as:

\begin{equation}
\begin{split}
    \overrightarrow{p}_{\rm SCA, 11}(x, y, \lambda)
    = 
    100 ( \overrightarrow{p}_{\rm grism}(x, y, \lambda) - \overrightarrow{p}_{\rm center}) + \frac{{\rm SCA}_{\rm width}}{2}
\end{split}
\end{equation}
\rst\ has 10 \micron\ pixel pitch (1 pixel per 10 \micron). We convert our vector difference from [mm] to pixel units with a multiplicative factor of 100. We then add half the SCA length to translate the origin in pixel scale from the middle of the detector to the bottom left of the detector pixel grid. We use the following expression for calculating $\overrightarrow{p}_{\rm SCA}$ for the (0,0) and (2,2) off orders:

\begin{equation}
\begin{split}
    \overrightarrow{p}_{\rm SCA, off}(x, y, \lambda) = 100 ( \overrightarrow{p}_{\rm grism}(x, y, 1.55) - \overrightarrow{p}_{\rm center}) \\
    + {\rm off}_{\rm 11, SCA}  + {\rm disp}_{\rm SCA} 10^4 (\lambda - 1.55) + \frac{\rm SCA_{\rm width}}{2} - 0.5
\end{split}
\label{eq:off-pos}
\end{equation}

Here ${\rm off}_{\rm 11, SCA}$ is the pixel offset from the off order to the (1,1) order at 1.55\micron, and ${\rm disp}_{\rm SCA}$ is the x- and y-dispersion of the off order. Both are unique to the SCA being simulated. 

Once the SCA pixel position is calculated, we only consider the points where $p_{\rm SCA, x} \& p_{\rm SCA, y} \in [0, 4096)$ for all evaluated orders. A source pixel's flux is entirely assigned to the nearest SCA pixel that it lands on, effectively utilizing a nearest neighbor approach when assigning flux. Because the image cube is oversampled by $3.7\times$ spatially and $11\times$ spectrally, relative to the final resolution of a simulated WFI image, this nearest-neighbor approach introduces only modest pixelation noise.

\subsection{Detector flux conversion}
\label{sec:grism:flux-assign}
Now knowing where a source pixel lands on the detector array, we are able to find how much of a source pixel's flux contributes to the grism pixel. We express this as following:

 \begin{equation}
    F_{\rm grism, px}(\overrightarrow{p}_{\rm SCA}, \lambda) = {10^{-17}}  f_{\rm px}(\lambda)  R(\overrightarrow{p}_{\rm SCA}, \lambda) \lambda_{\rm step}
\end{equation}

The flux density of the source pixel at the evaluated wavelength is first re-scaled to units of erg / s / cm$^2$ / \angstrom. This is then followed by converting the source flux into counts / s / \angstrom\ by the grism response function $R$ which has both positional ($\overrightarrow{p}_{\rm SCA}$) and wavelength ($\lambda$) dependence. We lastly obtained our desired units for the source pixel's grism flux (counts / s) by multiplying the result by the wavelength step.

The grism response function, $R$, has the following piecewise representation: 

\begin{equation}
\label{eq:response}
  R =
    \begin{cases}
      C_{(i,j)} R_{\rm grism}(\lambda)R_{\rm blue}(\overrightarrow{p}_{\rm SCA}, \lambda) & \text{$\lambda [\micron] \leq 1.004 $}\\
      C_{(i,j)} R_{\rm grism}(\lambda) & \text{$1.004 < \lambda [\micron] < 1.870$}\\
      C_{(i,j)} R_{\rm grism}(\lambda)R_{\rm red}(\overrightarrow{p}_{\rm SCA}, \lambda) & \text{$1.870 \leq \lambda [\micron]$}
    \end{cases}       
\end{equation}

We start with the base response component of the of the optical element, $R_{\rm grism}$, which varies with wavelength. The positional dependence comes into play when evaluating the response function towards the blue- or red-edge of the wavelength array. The response is scaled from 0 to 1 in the blue-edge regime ($1.000 < \lambda  \ [\micron] < 1.004 $) and from 1 to 0 in the red-edge regime ($1.87 < \lambda  \ [\micron] < 1.93 $). This factor is dependent on where the source pixel's photons land on the detector array.

\begin{equation}
    \begin{aligned}
    \lim_{\lambda \to 1.004 } R_{\rm blue}(\overrightarrow{p}_{\rm SCA}, \lambda) = 1 & &  \text{$\lambda [\micron] \leq 1.004 $}\\
    \lim_{\lambda \to 1.93} R_{\rm red}(\overrightarrow{p}_{\rm SCA}, \lambda) = 0 & & \text{$1.870 \leq \lambda [\micron] $}
    \end{aligned}       
\end{equation}

Initial modeling of this positional dependence is for a grid of the WFI consisting of 5 x 5 mm$^2$ areas (patches consisting of 500 x 500 WFI pixels$^2$) for wavelength steps of 5 \angstrom\ as see in Fig. \ref{fig:red-and-blue-edges}. When examining the blue-edge evolution, we see that the center of the WFI has a weaker response than the outer edges. The reverse is true when looking at the red-edge evolution - the center has a stronger response than the outer edges. In order to model this effect at the WFI pixel level, We perform 2D interpolation in order to finely sample this response per SCA pixel and our chosen wavelength step. Once converted, we assign the calculated flux for whichever order to the SCA pixel as determined from eq. 8.  We last apply a wavelength-dependent conversion factor $C_{(i,j)}$ for converting the main (1, 1) order response to that of the off order's: 

\begin{equation}
\label{eq:convert-response}
  C_{(i,j)} =
    \begin{cases}
      C_{(0,0)}(\lambda) & \text{order = (0,0)}\\
      C_{(1,1)}=1 & \text{order = (1,1)}\\
      C_{(2,2)}(\lambda) & \text{order = (2,2)}
    \end{cases}       
\end{equation}

\begin{figure}
	\includegraphics[width=\columnwidth]{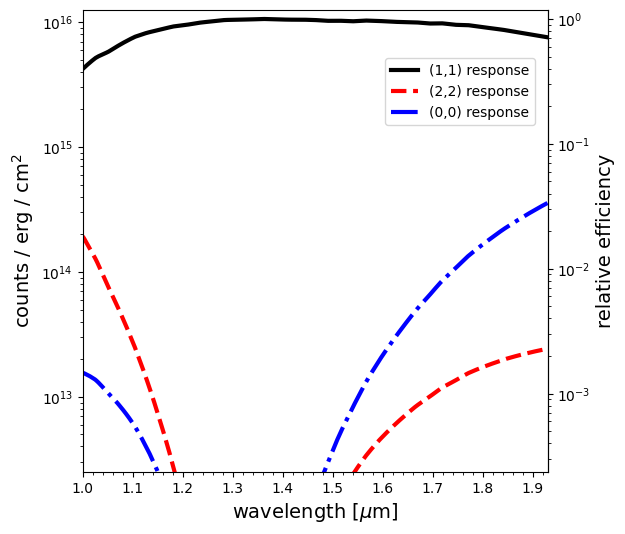}
    \caption{The grism's wavelength dependent response function $C_{(i, j)}R_{\rm grism}(\lambda)$ for the (0,0) (blue), (1,1) (black), and (2,2) (red) orders as expressed in eq. \ref{eq:response} \& \ref{eq:convert-response}. The relative efficiency is measured w.r.t. the maximum (1,1) response value. The positional based blue- and red-edge cutoffs are not included.}
    \label{fig:response-function}
\end{figure}

The converted wavelength dependent response function for all three orders in can be seen in Fig. \ref{fig:response-function}. We observe that the (2,2) order response is stronger than the (0,0) at the blue edge and the reverse at the red edge of the wavelength grid. 

\subsection{Source Injection}
\label{sec:grism:source-injection}
Source injection allows a user to inject a number of custom (usually unrepresented) source into the grism foreground simulations. The source injection pipeline requires three inputs - a configuration file, a morphology image stamp, and an observed frame SED. An optional segmentation map stamp for the image stamp can also be submitted, however one can be created on the fly by using all positive non-zero values in the image stamp. 

The image stamp is first pre-processed by setting negative flux values to 0, followed by a normalization such that the sum of the stamp's pixel values is unity, effectively calculating each pixel's fractional contribution as done before in \ref{sec:grism:data-cube}. Two blank canvases that are the same shape and pixel resolution as the science image are created for stamping sources onto. We also use the same WCS from the science image for the canvases to maintain consistency. A 2D array of pixel positions is also created to describe where the center of the stamp will be pressed on the canvas. The user can assign the stamping boundary in the parameter file if they want to populate the field in a certain area.    

The stamping loop is straightforward. The loop iterates through the pixel position array, the ID of the current object being the loop index. The object's center position is pulled from the position array and is added to the image stamp's pixel positions. The values of image stamp are thus stamped at the new pixel position on the image canvas. The segmentation stamp's values are multiplied by the ID before similarly being stamped on the same location on the segmentation canvas. This process is repeated for the number of objects specified, and the science and segmentation canvases are then saved as compressed FITS files once the process is complete. 

The SED is pre-computed to both set negative flux density values to 0 and to have a wavelength range of 1 - 1.93 $\micron$ at a given $\lambda_{\rm step}$ as described in Sec. \ref{sec:grism:data-cube}. A rest-frame SED is shifted to the correct observed frame based on the specified redshift in the parameter file. An SED library is concurrently created during the stamp loop, where the SED is repeatedly assigned to each ID in a dictionary for all objects in the loop.

\begin{figure*}
    \includegraphics[width=\textwidth]{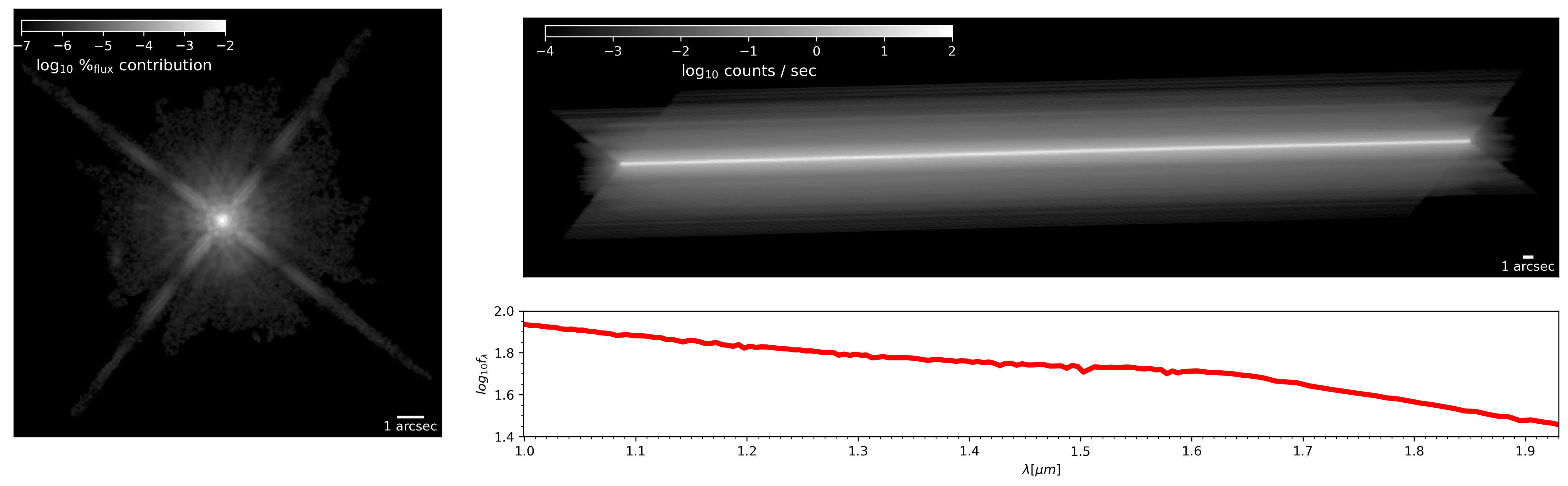}
    \caption{An example of a star injected into the middle of SCA5. The left panel is the image stamp with an area of 550 x 550 px$^2$ (16.5" x 16.5"), taken directly from \hst\ CANDELS F125W imaging data (COSMOS ID - 15101). The top-right is the simulated grism (1,1) order cutout of 860 x 100 px$^2$ (94" x 11"). The bottom-right is the input SED from 1-1.93$\micron$ at with $\lambda_{\rm step} = 1$\angstrom.}
    \label{fig:star-inject}

    \includegraphics[width=\textwidth]{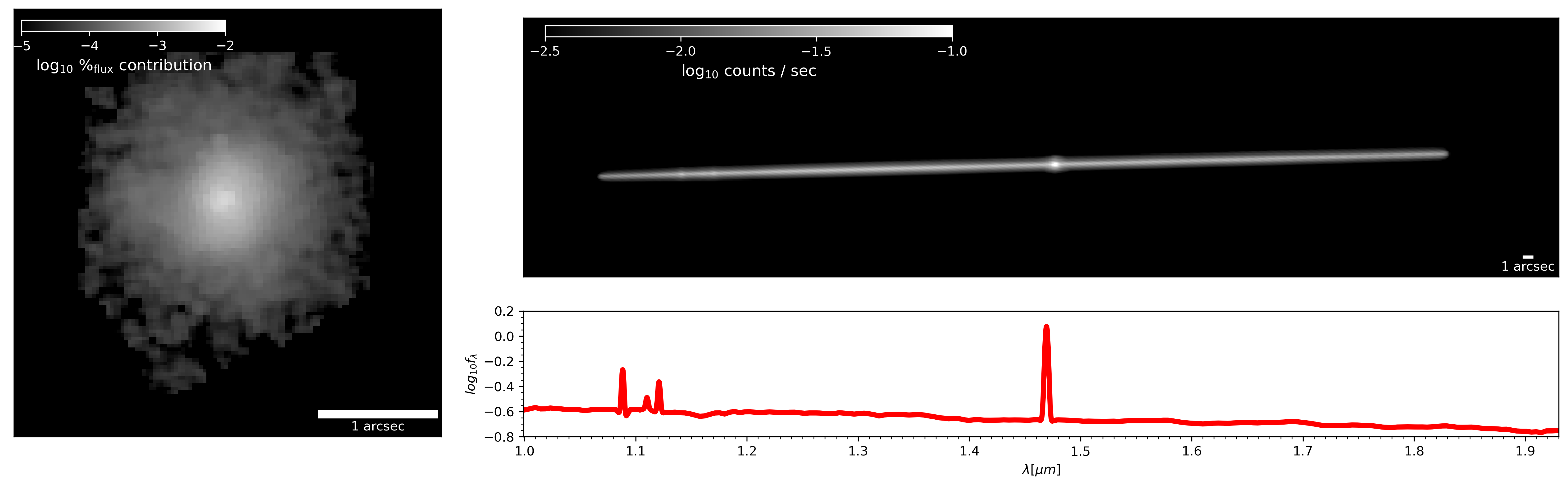}
    \caption{An example of an emission line galaxy (ELG) injected into the middle of SCA5 with H$\alpha$, H$\beta$, and [OIII] emission lines. The properties of the galaxy are the following: COSMOS ID - 15485, $z \sim 1.27$, ${\rm log}_{10} {\rm M}_{*} = 10.74 \rm[ M_\odot]$ \citep{3d-hst:skelton, suess:2019}. Panels are similarly described in Fig. \ref{fig:star-inject}. The stamp has an image area of 120 x 120 px$^2$ (3.6" x 3.6").}
    \label{fig:elg-inject}

    \includegraphics[width=\textwidth]{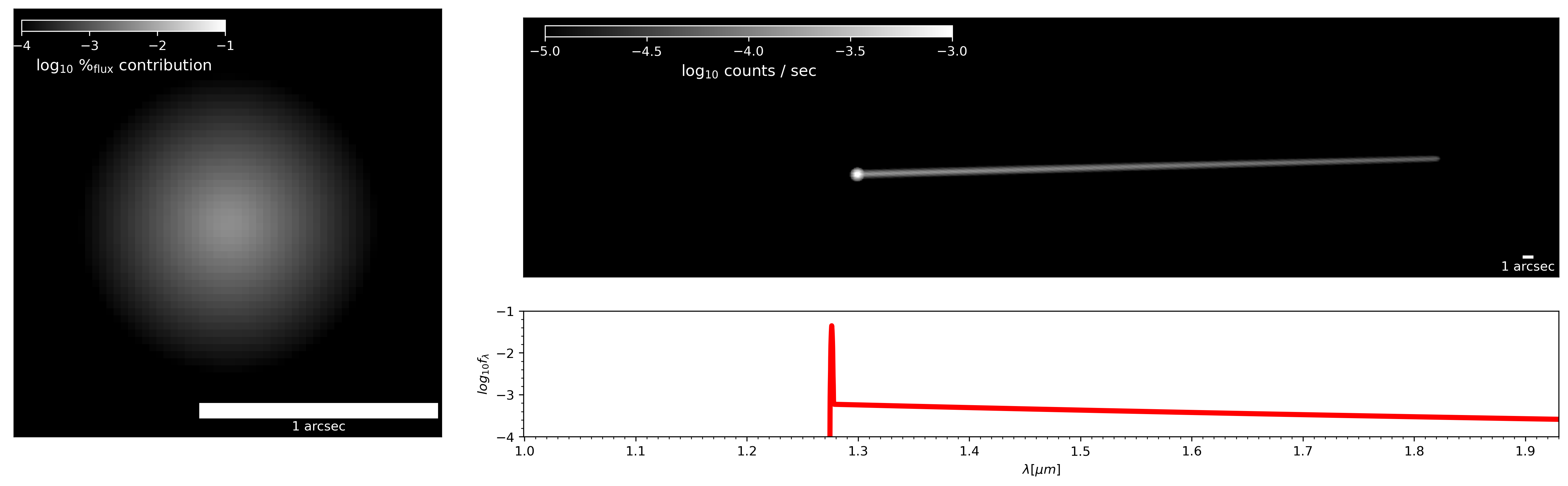}
    \caption{An example of a $z = 9.5$ Ly$\alpha$ emitter (LAE) injected into the middle of SCA5 using the process described in Sec. \ref{sub:lae}. Panels are similarly described in Fig. \ref{fig:star-inject}. The stamp has an image area of 50 x 50 px$^2$ (1.5" x 1.5").}
    \label{fig:lae-inject}
\end{figure*}

The three resulting products then can be used as input in the data cube constructor (as outlined in subsection \ref{sec:grism:data-cube}) and the grism simulator (Sec. \ref{sec:grism:polynomial} - \ref{sec:grism:flux-assign}). The simulation produces a grism image of just the injected objects, which can be linearly combined with the foreground simulation to produce a realistic test scene for object recovery. This is because of each object's spectrum and morphology is considered independent from all others, where the only interaction is contamination in the grism step when the flux from two different objects lands on the same detector pixel. Figs. \ref{fig:star-inject} and \ref{fig:elg-inject} both showcase the injected inputs and output for a bright, broadband star ($m_{\rm F160W} = $, spectral type ) and an emission line galaxy (ELG; $m_{\rm F160W} = $) respectively. We note that the star is taken directly from \hst\ imaging so the injection result inherits the \hst\ PSF. The source still demonstrates the contaminating effects of bright diffraction spikes and their relative contribution to the scene. The injected ELG, whose stamp was also directly obtained from \hst\ imaging, showcases emission line spikes in the 2-D spectra that are present in the input 1-D spectrum. Objects like the ELG are ideal for testing source recovery software.

\subsubsection{Ly-$\alpha$ emitters}
\label{sub:lae}
The \rst\ grism has the capability to detect Lyman-$\alpha$ emission at a redshift range of $ 7.3 \lesssim  z \lesssim 14.9$, making it a great instrument for constraining properties of reionization. We inject custom LAEs via our source injection pipeline with varied parameters to compensate for the absence of these objects in the base science image. Our creation and treatment of mock LAEs follow the same treatment outlined in \cite{wold:2023}.

Our initial LAE morphology is generated by a S\'ersic profile with a S\'ersic index of n = 1 and a half-light radius r = 0.25 kpc to emulate the compact morphology seen in the local universe \citep{kim:2021}. The science stamp is then converted from the original 10 mas / pixel resolution to the same of the science image (30 mas / pixel) by summing up 3 x 3 pixel areas. The stamp is normalized to unity, and we apply a radial mask on the science image to create a segmentation stamp of the LAE. Both the science and segmentation stamps are used for all injected LAEs regardless of their 'physical' properties.  

We model our LAE SEDs via a simplified model of a Gaussian peak with an attenuated continuum. First, the properties of each LAE is determined by the following sampling criteria:

\begin{itemize}
    \item $\frac{d \rm N}{d \rm L} \propto {\rm L}^{\alpha}$, where $\alpha = -2.5$, min(L$_{\rm Ly\alpha}$) = 10$^{42.6}$ erg s$^{-1}$
    \item $\frac{d \rm N}{d\rm EW} \propto {\rm E}^{-{\rm EW} / l_{\rm scale}}$, where $l_{\rm scale}$ = 100 \angstrom\ , min(EW) = 10 \angstrom 
    \item $7.25 < z < 10.5$, where z$_{\rm step}$ = 0.25 
\end{itemize}

Our choice $\alpha = -2.5$ is based off of the $z \sim 7$ observations of the Ly$\alpha$ luminosity function from the LAGER survey (Lyman Alpha Galaxies in the Epoch of Reionization) \citep{zheng:2017, hu:2019, wold:2022} and z = 5.5 and 6.6 functions from the SILVERRUSH program (Systematic Identification of LAEs for Visible Exploration and Reionization Research Using Subaru HSC) \citep{konno:2018}. The equivalent width scale length is consistent with spectroscopic results taken with MUSE (Multi Unit Spectroscopic Explorer) \citep{hashimoto:2017}. The sampling determines the redshift, Ly$\alpha$ line flux (cgs), and continuum magnitude (AB mag) of each object. 

 Our base wavelength array is shifted to the observed frame based on the LAE's redshift, followed by calculating the observed continuum flux by evaluating continuum magnitude in the F160W filter with a zero point of -48.6. Our mock base continuum follows the following power law representation: $f_{\rm \lambda} \propto \lambda^{-2}$. The base continuum is then attenuated by the IGM prescriptions described in \cite{inoue:2014} which models the transmission functions for the Ly$\alpha$ forest and damped Ly$\alpha$ systems. The continuum is subsequently scaled by the ratio of the continuum flux and the the value of the attenuated continuum at the F160W pivot wavelength. Our emission line is modeled via a gaussian that is scaled by the Ly$\alpha$ line flux. The emission line is added on top of the continuum, and we re-sample our new LAE SED within the grism's wavelength range of $1 - 1.93 \micron$ with $\lambda_{\rm step} = 1$ \angstrom\ as our final step. 

We assign the positions of our LAEs to lie in the same 0.025 deg$^2$ cutout region from the COSMOS field as seen in Fig. \ref{fig:sci-img-cutout} which avoids galaxies existing outside of the foreground area. A total of 5000 synthetic LAEs are created and injected into the image. An example of an injected LAE is seen in Fig. \ref{fig:lae-inject}. All emission to left the emission line has been attenuated. The predicted 2D spectra shows in isolation that the LAE will appear as a bright emission feature followed by slowly dimming fainter tail.

\subsection{Post-processing effects}
\label{sec:grism:additional}
We perform a final series of operations to produce our realistic images. First, we linearly combine our foreground image with our injected LAE image to include the sources in a final scene. We then model photon noise, which is expected to be the dominant noise source for faint objects in deep \rst\ grism observations. We assume a sky background of 0.8 counts / s, which we  uniformly add to each detector pixel.  We then multiply by the simulated exposure, typically taken to be $t_{\rm exp} = 10^4$ s, which shifts our images to units of counts. We add Poisson noise to the image model based on these counts, to simulate photon counting noise.  Finally, we divide by $t_{\rm exp}$ to return back to base units of counts / s. The result is noisy image where 'empty' patches (areas without data) have a standard deviation $\sigma \sim  \sqrt{0.8 t_{\rm exp}} / t_{\rm exp}$. Last, We zero out the flux for all SCA reference pixels to only include active pixels in the final image result. 

\subsection{Pipeline parameter files}
\label{sec:grism:parameters}
\begin{figure}
	\includegraphics[width=\columnwidth]{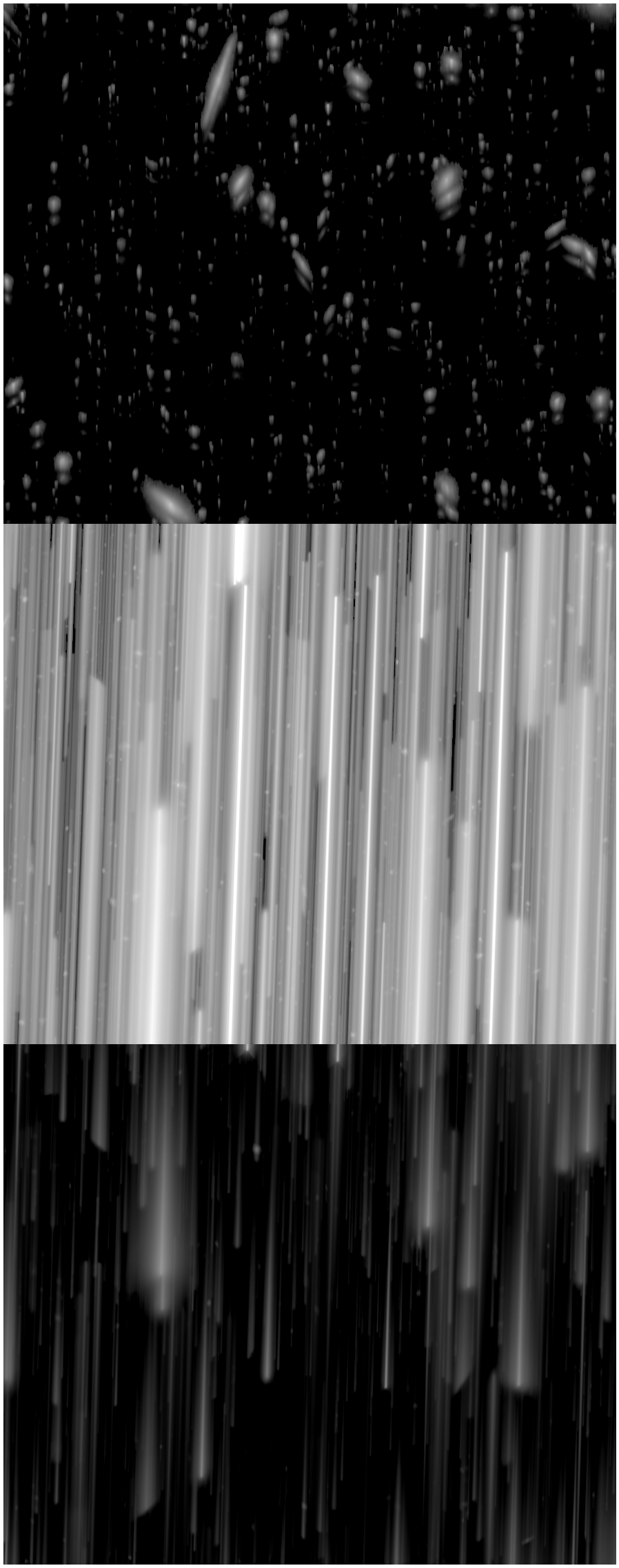}
    \caption{Scaled noiseless images of the (top) (0,0), (middle) (1,1), and (bottom) (2,2) orders. Each frame shows a the same region of 110 by 93.5 arcseconds (1000 by 850 pixels at a resolution of 11 mas / pixel).}
    \label{fig:orders}
\end{figure}

Running the simulation requires a number of user-defined inputs that can be adjusted or toggled via input parameter files. To describe some, the datacube parameters include specifying the total wavelength range to generate the cube, what $\lambda_{\rm step}$ interval to generate the datacube, and cropping dimensions of the source and segmentation map images. The grism simulation parameters include flags to toggle on and off the main science order and each off order (as demonstrated in Fig. \ref{fig:orders}), as well as red-/blue-edge cutoffs. The grism parameters also define which detector to simulate, dithering, as well as simulation rotation angle. Last, the injection parameter file allows a user to define how many of the desired object to inject into the field and within what area. All aforementioned parameter sets require defined paths to relevant input images and directories.  While separate parameter files can be created for each of the described processes, a single parameter file will work as well so long as each header in the file is defined correctly. 

\section{Discussion}
\label{sec:discussion}

\subsection{Comparisons to other simulations}
There is a growing literature volume dedicated to \rst\ preparatory science, a few of those dedicated to creating mock grism foreground images. One such example is from \cite{wang:2022}, who similarly developed a slitless spectroscopy pipeline for mock observations for the \rst\ High-Latitude Spectroscopy Survey (HLSS). These simulations were generated using the software package aXeSIM \citep{axe:kummel}, originally developed to create mock grism and direct images for \hst. Note that aXeSIM disperses an object's light along the x-axis, so a position angle of $90^{\circ}$ is used to match \rst's y-axis dispersion. The primary input to the simulation is a mock lightcone produced by Galacticus \citep{benson:2012}, a semi-analytic model of galaxy formation, specifically a galaxy catalog consisting of $0 < z < 3$ objects with an \rst\ H-band (F158) magnitude of 28. Notable differences, barring software and pipeline-based ones, include that we are directly modeling sky-to-detector distortions (including \rst's y-axis dispersion), modeling objects out to $z \sim 5$, and have produced a more complete suite of roll angles and dithers, albeit for a smaller imaging area. 

\begin{figure*}
    \includegraphics[width=\textwidth]{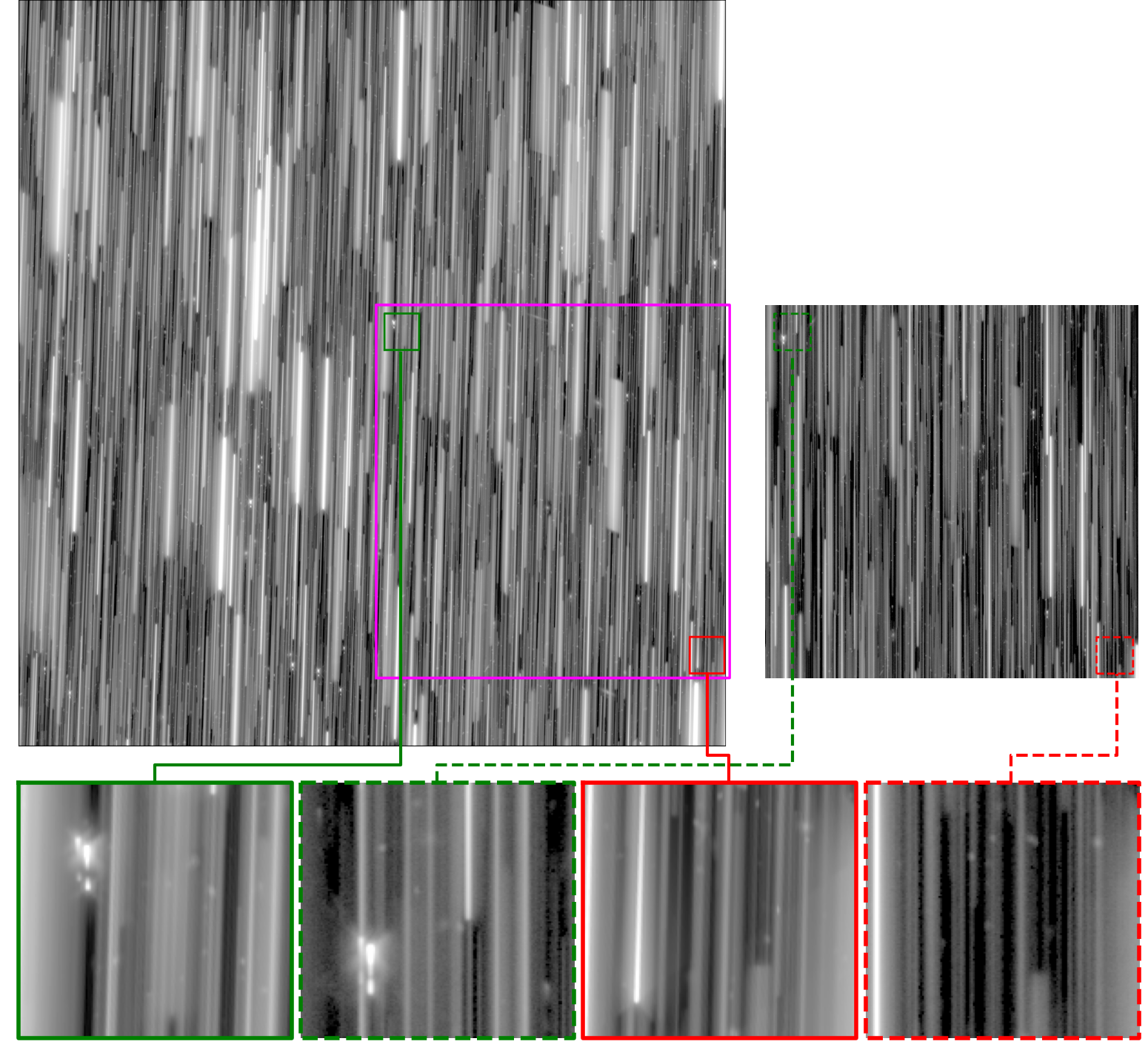}
    \caption{A comparison of \texttt{ESpRESSO} to other simulation results. Top-left: COSMOS grism simulation produced using \texttt{ESpRESSO} at a position angle at $90 \deg$. Top-right: COSMOS grism simulation produced using aXeSIM. Bottom row: 22 x 22 arcsec$^2$ (200 x 200 px$^2$) cutouts comparing best matched cutouts from both simulations from the top-left (green) and bottom-right (red) corners. The area enclosed in magenta is the best matched area to the aXeSIM simulation from \citet{wold:2023}.}
    \label{fig:sim-comparisons}
\end{figure*}

The foreground simulations produced in \cite{wold:2023}, also created with aXeSIM, were created to test the source recovery tool, CUBGRISM, which was originally constructed to reduce and extract sources from Galaxy Evolution Explorer (\textit{GALEX}) grism data \citep{wold:2014, wold:2017}. These aXeSIM simulations modeled 1/4 of the SCA5 area using similar methodology to the \texttt{ESpRESSO} simulations, e.g. COSMOS field input images, 3D-HST spectra, modeling the same off orders and the wavelength-dependent response, etc. This allows us to make more direct comparisons with aXeSIM-based simulations. Fig. \ref{fig:sim-comparisons} demonstrates visual differences between the \texttt{ESpRESSO} and Wold aXeSIM simulations both at the macro and micro levels. Both simulation images are similarly scaled, with the best-matched imaging area highlighted. Finding an exact match is difficult due to different choices in field center, which then affects the dispersion solution per object, as well as different undeviated wavelengths ($0.95\ \mu m$ for the (1,1) order in aXeSIM, $1.55\ \mu m$ in \texttt{ESpRESSO}) and input image resolutions (0.06 vs. 0.03 mas). We first examine the macro-scale, where we observe differences in the 2D spectra flux that are presumed to originate from the same sources. This is due to aXeSIM normalizing the input SEDs to match a catalogued filter profile's flux, specifically F160W, while \texttt{ESpRESSO} directly uses the input SED with no modification. Micro level differences are highlighted in the 200 x 200 px$^2$ (22 x 22 arcsec$^2$) patches at the top-left and bottom-right of the  best matched areas. The dispersion solution for all three orders is noticeably different between the \texttt{ESpRESSO} and aXeSIM simulations. This is due to a slight deviation of the (1, 1) trace from the y-axis as we move away from field center which is accounted for in \texttt{ESpRESSO}. This angular deviation is then propagated to the (0, 0) and (2, 2) dispersion solutions based on Eq. \ref{eq:off-pos}. 

\subsection{Off order or emission line pair?}
Because zero-order light from slitless grism data is spatially compact, it can be mistaken for an emission line in a first order spectrum under some conditions.   Here we examine the likelihood of such confusion for the {\it Roman} grism.  For {\it Roman}, the (0,0) order has a small but nonzero dispersion, which separates the blue and red light into a double-peaked structure with a minimum where the grism reaches its peak (1,1) order efficiency. This is clearly seen in the lower left panels of fig.~\ref{fig:sim-comparisons}.   

We find that the separation between the two peaks is (on average) 29 pixels, and the blue:red flux ratio is 1:9, as expected from the grism response functions shown in Fig.~\ref{fig:response-function}.
The (1,1) order has a dispersion of $11 \ \angstrom$ per pixel, so the peaks of the (0,0) could be misinterpreted as an emission line pair (L$_1$, L$_2$) with an observer-frame wavelength separation of $\Delta \lambda = 319 \ \angstrom$.   

For such a pair to correspond to a physical doublet would require $\Delta \lambda_{\rm rest} \times  (1+z_{\rm pair})= \Delta \lambda_{\rm obs}$.
Rearranging, we can find the $z_{\rm pair}$ required for a particular line pair to have spacing matching a (0,0) structure:
\begin{equation}
    z_{\rm pair} = \frac{\Delta \lambda_{\rm obs}}{\Delta \lambda_{\rm rest}} - 1
    = \frac{319 \ {\rm \angstrom }}{\lambda_{\rm 2,rest} - \lambda_{\rm 1,rest}} - 1
\end{equation}

\begin{table}
    \centering
    \begin{tabular}{c|c|c}
        L$_1$ & L$_2$ & $z_{\rm pair}$ \\
        \hline
        \sII $\lambda6732\angstrom$ & \nII $\lambda6549\angstrom$ &  0.71 \\
        \sII $\lambda6718\angstrom$ & \nII $\lambda6549\angstrom$  & 0.86 \\
        \sII $\lambda6732\angstrom$ & H$\alpha$  & 0.87 \\
        \sII $\lambda6718\angstrom$ & H$\alpha$ & 1.04 \\
        \sII $\lambda6732\angstrom$ & \nII $\lambda6585\angstrom$  & 1.14 \\
        \oIII $\lambda5008\angstrom$ & H$\beta$ & 1.19 \\   
        \sII $\lambda6718\angstrom$ & \nII $\lambda6585\angstrom$ & 1.36 \\
        \oIII $\lambda4960\angstrom$ & H$\beta$ & 2.27 \\
        \nV & Ly$\alpha$ & 11.47 \\
    \end{tabular}
    \caption{Table of emission line pairs that have a separation of $319 \angstrom$ at the redshift $z_{\rm pair}$. $L_1$ is the blue-r line, $L_2$ the redder. Table is sorted on $z_{\rm pair}$.}
    \label{tab:line_pairs}
\end{table}

We calculate the possible pairs for galaxy and quasar emission lines, based on those listed in the Sloan Digital Sky Survey Data Release 6 (SDSS DR6) documentation \citep{mccarthy:2008:sdss-vi}. The possible line pairs and the redshifts with potential for confusion are shown in Table \ref{tab:line_pairs}. Most of these can be ruled out using considerations besides wavelength separation.   

The first group of lines pairs fall under the \sII\ -- L$_2$ umbrella. The majority of these line pairs can be ruled out due to the presence of other, nearby emission lines expected in the grism's wavelength range. H$\alpha$, for example, lies in between the \nII doublet, and is brighter than either line, ruling out any \sII\ -- \nII\ doublet combination. The \sII\ doublet -- H$\alpha$ combinations are implausible because H$\alpha$ is almost always brighter than the \sII\ doublet, while the (0,0) artifacts are brighter at the red peak.  Similar arguments can be used to rule out most of the remaining pairs, e.g., \oIII\ -- H$\beta$ will usually appear as a triplet in the resolved range. 

A more plausible scenario for confusion arises when the (0,0) order has a low signal-to-noise ratio, so that the blue peak is below the detection threshold.  Direct imaging of the desired targets can be and should be used to mitigate this possibility.   Because the (0,0) image contains a modest fraction of the total light, even fairly shallow direct images containing the source of the (0,0) light should detect that source, and \texttt{ESpRESSO} modeling of the full scene should predict the location of the contaminating (0,0) order source.

\subsection{Investigating scene crowding}

\begin{figure}
    \includegraphics[width=\columnwidth]{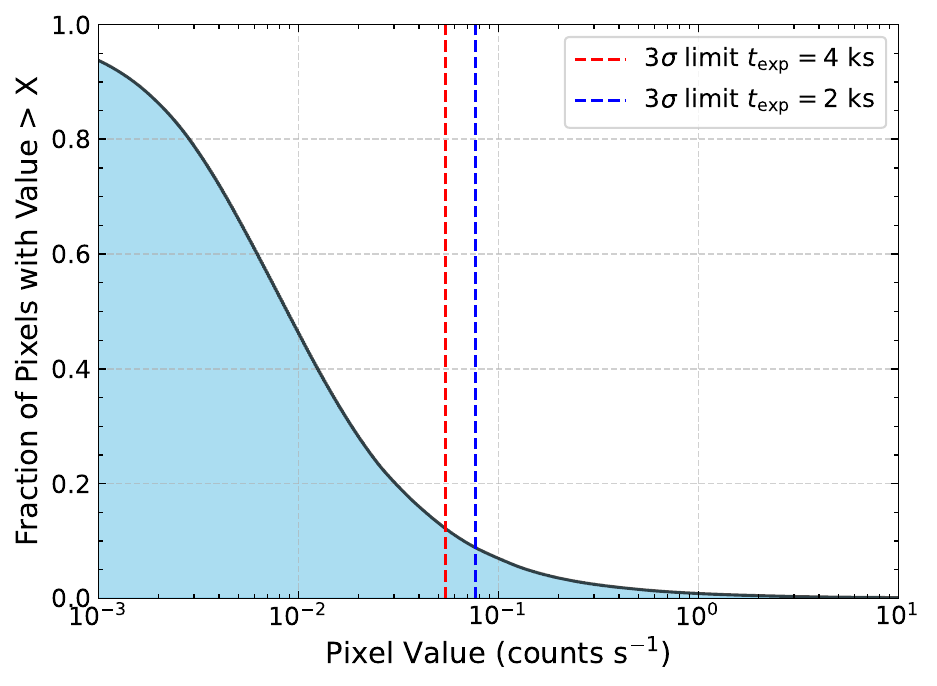}
    \caption{For our simulated grism extra-galactic scene, we show the cumulative fraction of pixels with flux above a count per second threshold.  Assuming a background level of $1.3$ counts s$^{-1}$ pixel$^{-1}$, we show the $3\sigma$ sky-background limit for typical High Latitude Wide Area survey exposure times  (dashed vertical lines).  Approximately $10\%$ of the pixels will have a significantly elevated background level due to foreground objects.}
    \label{fig:pixelfrac}
\end{figure}

Fig. \ref{fig:sim-comparisons} gives the impression that the extragalactic foreground covers a large percentage of the simulated grism image. This is because this is a noiseless simulation made from deep images. The real question is what fraction of pixels experience a higher photon count due to crowding. We quantify this in Fig. \ref{fig:pixelfrac} by showing the cumulative fraction of pixels with flux above a count per second threshold. The depth of the grism observations stated in the Roman WFI technical report\footnote{ \url{https://roman.gsfc.nasa.gov/science/WFI_technical.html}} assumes a sky-limited regime where the sky background is $2\times$ minimum zodiacal, $1.3$ counts s$^{-1}$ pixel$^{-1}$.  Given this sky level and the typical exposure times of the High Latitude Wide Area survey, we find that foreground sources will significantly boost the background level for about $\sim10\%$ of the grism survey.  This emphasizes the need for reduction algorithms that can de-conflict overlapping sources.  Regardless, we conclude that the assumption of a sky-limited background is valid for $\sim90\%$ of grism field of view. 

\subsection{Caveats and limitations of \texttt{ESpRESSO}}
While we have made design choices to produce simulations of high fidelity, both in instrument parameters and the astrophysical scene, some limitations remain due to incomplete knowledge at the time of development of the code. It is important to remember and consider the synthetic images when utilizing them in upcoming source recovery tools development. We highlight both input and simulation-based limitations.

(1) The grism parameters:  such as dispersion, throughput, and field distortions are taken from the grism model as designed, and not as it will perform in space. These parameters can, however, be updated in \texttt{ESpRESSO} with actual parameters derived from performance after launch. The fidelity should be good enough for planning purposes. 

(2) The deep extragalactic scene: We take deep imaging in CANDELS by the Hubble as the starting point of the simulations. This is done so we get realistic shapes of galaxies and also their clustering and overlap. The CANDELS image is somewhat deeper than the High Latitude Wide Area Survey (HLWAS)\footnote{ \url{https://science.nasa.gov/mission/roman-space-telescope/high-latitude-wide-area-survey/}} is expected to be in similar bandpasses, and the area covered on sky is much smaller. The spatial resolution of Hubble is also very close to what we expect from Roman, even though the detailed shape of the PSF is expected to be different. The Roman Point Spread Function is wavelength dependent and differs in the number of diffraction spikes from the Hubble PSF. We chose to use the CANDELS images directly without deconvolving with the \hst\ PSF and then convolving with  \rst\ PSF.  This choice was made for computational simplicity, and is unlikely to affect anything except bright stars and their immediate neighbors.  The other limitation arising from this approach is the field size that we are able to simulate with the grism. Our current simulations have a total area coverage of $\sim$ 2-3 \rst\ SCAs per position angle, nowhere near the full detector array. \texttt{ESpRESSO} is flexible enough to apply to larger fields of view as and when they become available.

(3) The  EAZY SED library, which is primarily useful in providing easily accessible 1D spectra across a wide wavelength range from infrared to ultraviolet, is built on a limited set of templates that do not represent the full variety of objects present in our scene. As mentioned in Sec.~\ref{sec:data:3dhst&eazy}, there are no templates for re-constructing the SEDs of bright point sources (e.g. stars), which are important for properly modeling foreground scene contamination.
Our mitigation was to use approximate SEDs for these objects, selected based on their colors, as outlined in section~\ref{sec:sed_correction}.
The templates also do not cover emission and absorption from rarer object classes, such as multiple types of AGN and QSOs.  This can be addressed using \texttt{ESpRESSO} tools by injecting the additional source types into our scene using custom SEDs.

(4) We assume that the SED flux at a given wavelength is applied to entire object when constructing the image datacube. While this is a sufficient simplification, it is not wholly grounded in truth as it implies that a specific emission line is being produced by the entire galaxy. This results the entire galaxy brightening at observed emission line wavelengths.  Figs. \ref{fig:elg-inject} and \ref{fig:lae-inject} show this, as the input emission line brightens the entire object's morphology in the simulated 2D spectrum. A more realistic approach would be modeling a galaxy with a broadband SED while modeling sub-regions that producing the specific spectral emission lines. The galaxy and sub-region grism results could then be co-added, a similar process to source injection. This affects only a small fraction of galaxies that are relatively large on the sky, with distinct HII regions. 

(5) We are also currently modeling a subset of the potential grism orders, in particular the three orders (0:0, 1:1, 2:2) that are in focus and which produce the highest surface brightness contributions to the grism output. The other  orders are defocused and produce very low surface brightness contamination except in case of extremely bright stars.

\section{Conclusions}
\label{sec:conclusion}

We have presented \texttt{ESpRESSO}, a new slitless spectroscopy simulation code developed for \rst\ preparatory work.  We have shown both the flexibility in creating foreground images and that our simulated grism foregrounds can be modified to superimpose custom sources and exposure time noise, making them ideal for upcoming grism data analysis and extraction tools ahead of {\rst\/'s launch. Our key results are:

\begin{itemize}
    \item The (0,0) order images are unlikely to be mistaken for true emission line pairs even when they land on top of a spectrum.  \item Based on current core-community survey design, foreground contaminants will affect about 10\%  of the pixels. 
\end{itemize}

\section*{Acknowledgements}

The authors of this paper would like to thank Ami Choi, ChangHoon Hahn, Kartheik Iyer, Peter Melchior, Gregory Mosby, Lucia Perez, Andreea Petric, Rachel S. Somerville, and L. Y. Aaron Yung for helpful questions and discussion. 

The material is based upon work supported by NASA under award number 80GSFC21M0002. This work is based on observations taken by the 3D-HST Treasury Program (GO 12177 and 12328) with the NASA/ESA \hst\, which is operated by the Association of Universities for Research in Astronomy, Inc., under NASA contract NAS5-26555.  Partial support for this work was provided by the WFIRST Science Investigation Team contract NNG16PJ33C, `Studying Cosmic Dawn with WFIRST'.  Resources supporting this work were provided by the NASA High-End Computing (HEC) Program through the NASA Center for Climate Simulation (NCCS) at Goddard Space Flight Center.

The following software and packages were utilized in this work: Python 3 \citep{python3}, astropy \citep{astropy:i, astropy:ii, astropy:iii}, eazy-py \citep{eazy-py}, h5py \citep{h5py}, matplotlib \citep{matplotlib}, multiprocessing, numba \citep{numba}, numpy \citep{numpy}, pandas \citep{pandas}, and scipy \citep{scipy}, SAO DS9 \citep{sao:ds9}, Source Extractor \citep{sextractor:bertin&arnouts}
          }
\section*{Data Availability}

All grism simulation image products produced and used in this work will be made available to the public upon acceptance to the journal.



\bibliographystyle{mnras}
\bibliography{main} 




\appendix

\section{Supplemental Figures}

\begin{figure*}
    \includegraphics[width=0.98\textwidth]{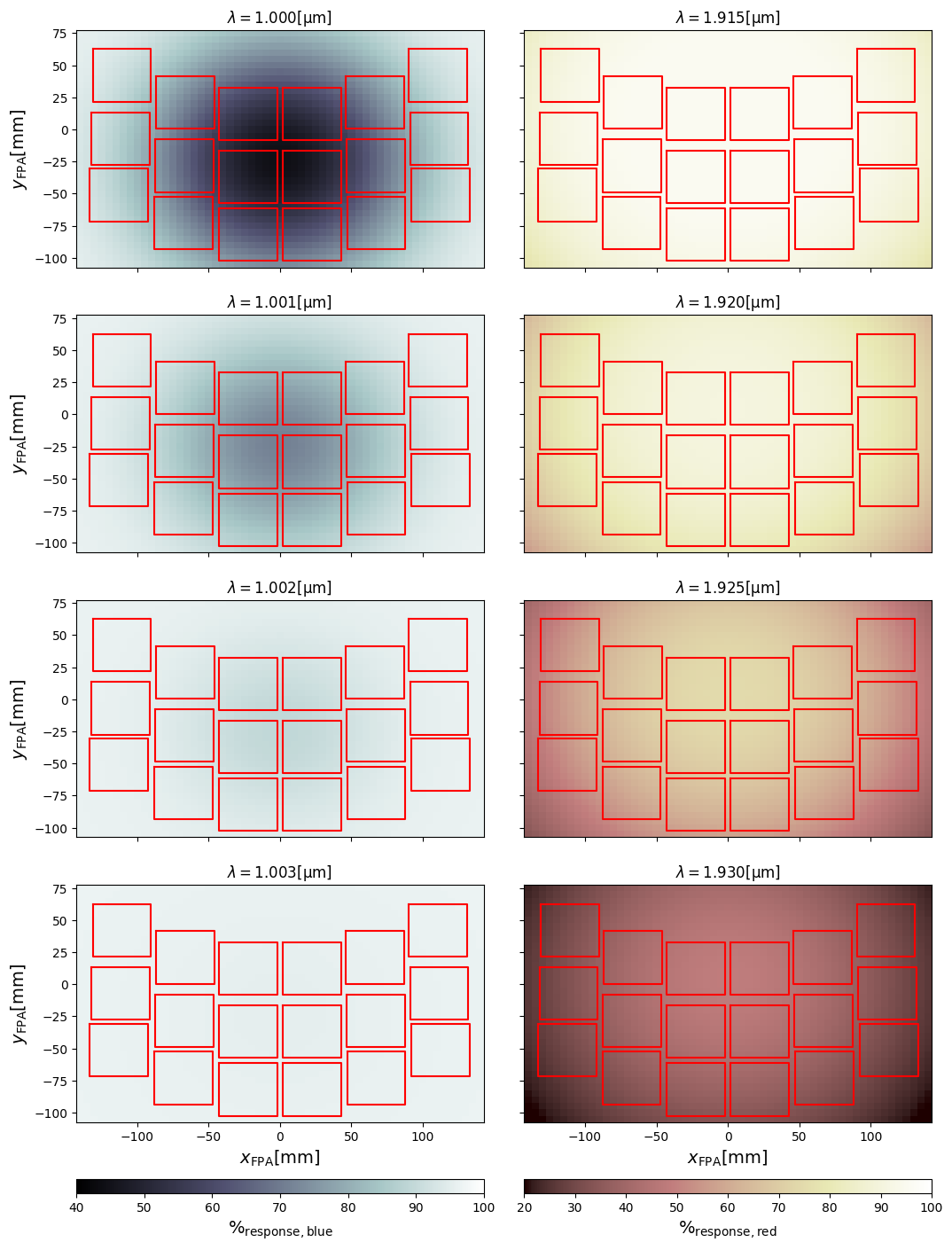}
    \caption{(left) Blue-edge (1.000 - 1.004 $\mu$m) cutoff response evolution. (right) Red-edge (1.87 - 1.93 $\mu$m) cutoff response evolution.}
    \label{fig:red-and-blue-edges}
\end{figure*}

In this appendix, we showcase additional figures related to \rst\/'s detector sensitivity. We showcase the red and blue wavelength edge cutoffs in Fig. \ref{fig:red-and-blue-edges} at select wavelengths and how the field's response varies with both position and wavelength. The heatmaps are the raw (non-interpolated) field dependent red- and blue-edge responses.  We see that as $\lambda \rightarrow 1.0 \ \micron$ that the center of the detector array has the least response while as $\lambda \rightarrow 1.93 \ \micron$ that the edges of the array have a dampened response. The interpolation and usage of the response is described in Sec. \ref{sec:grism:flux-assign}.


\bsp	
\label{lastpage}
\end{document}